# Visualising the Vertical Energetic Landscape in Organic Photovoltaics

Vincent Lami[a,b], Andreas Weu[a,b], Jiangbin Zhang[c], Yongsheng Chen[d], Zhuping Fei[e], Martin Heeney[e], Richard H. Friend[c] and Yana Vaynzof [a,b*]

a. Kirchhoff-Institut für Physik, Ruprecht-Karls-Universität Heidelberg, Im Neuenheimer Feld 227, 69120 Heidelberg, Germany.

b. Centre for Advanced Materials, Ruprecht-Karls-Universität Heidelberg, Im Neuenheimer Feld 225, 69120 Heidelberg, Germany.

c. Cavendish Laboratory, University of Cambridge, JJ Thomson Avenue, Cambridge CB3 0HE, United Kingdom

d. The Centre of Nanoscale Science and Technology, Key Laboratory of Functional Polymer Materials, State Key Laboratory and Institute of Elemento-Organic Chemistry, College of Chemistry, Nankai University, Tianjin, 300071, China

e. Department of Chemistry and Centre for Plastic Electronics, Imperial College London, London, SW7 2AZ, UK

Corresponding author: Yana Vaynzof, email: vaynzof@uni-heidelberg.de

Energy level diagrams in organic electronic devices play a crucial role in device performance and interpretation of device physics. In the case of organic solar cells, it has become routine to estimate the photovoltaic gap of the donor:acceptor blend using the energy values measured on the individual blend components, resulting in a poor agreement with the corresponding open-circuit voltage of the device. To address this issue, we developed a method that allows a direct visualisation of the vertical energetic landscape in the blend, obtained by combining ultraviolet photoemission spectroscopy and argon cluster etching. We investigate both model and high-performance photovoltaic systems and demonstrate that the resulting photovoltaic gaps are in close agreement with the measured CT energies and open-circuit voltages. Furthermore, we show that this method allows us to study the evolution of the energetic landscape upon environmental degradation, critically important for understanding degradation mechanisms and development of mitigation strategies.

*Lead contact: Yana Vaynzof, e-mail: vaynzof@uni-heidelberg.de*

## Introduction

In recent years, significant advances in the development of organic photovoltaic (OPV) devices resulted in power conversion efficiencies (PCEs) surpassing 15% and 17.3% for single junction and multi-junction cells, respectively.[1,2] One of the key device parameters that still introduces significant losses and should be optimised further is the open-circuit voltage ($V_{OC}$) of photovoltaic (PV) diodes. It is generally accepted that the $V_{OC}$ is associated with the 'photovoltaic gap', namely the energetic difference between the lowest unoccupied molecular orbital (LUMO) of the acceptor and the highest occupied molecular orbital (HOMO) of the donor at their interface.[3,4] This suggests that the first step for $V_{OC}$ optimisation and elucidation of energetic losses in the device must include an accurate determination of the energy levels of the donor and acceptor materials and most importantly their energetic alignment at the hetero-interfaces of the device active layer. This is particularly important for the new range of high efficiency photovoltaic systems with low driving energy offset between the energy levels of the donor and either the fullerene[5,6] or the non-fullerene acceptor,[7–11] as well as for novel ternary blend systems.[12,13] The energetic alignment of these systems remains poorly characterised, though studies of charge-transfer (CT) luminescence do provide important information.[14,15] These studies offer great insights into the loss mechanisms originating from the offset between the bandgap of the donor and the CT state at the heterojunction interface. For example, it was shown that the excess of energy in the initially generated excited state with respect to the CT ground state is of little significance, as most photoexcitations into the higher energy CT states relax within the CT manifold until reaching thermal equilibrium with the CT ground state prior to charge separation.[14] While the CT luminescence method has been utilised to show that some unavoidable non-radiative voltage losses, linked to electron-vibration coupling, may be present in organic systems that employ fullerenes,[15] recent works suggest that this may not be the case for systems with low energy offsets.[16,17] Despite the great insights available from this method, it has been shown to be sensitive to variations in active layer thickness and device architecture,[18] and ultimately does not provide information about the energetic landscape in organic devices. Currently, measurement techniques that offer this information are very limited. While reporting of energy levels is ubiquitous in studies investigating OPV devices,[19,20] they often combine energy values obtained by different techniques, including density functional theory calculations,[21,22] cyclic voltammetry (CV),[22,23] differential pulsed voltammetry,[21] Kelvin probe,[24,25] and photoemission spectroscopy,[22,26–28] resulting in a large scatter of reported energies, even for the same materials. The impact of polarization effects,[29] is also often not taken into account. Additionally, energy level diagrams often neglect to account for interfacial effects such as the formation of dipoles or band bending.[29–31] Several reports attempted to obtain accurate information about the energetic landscape of the device active layer. It has been, for example, approximated by performing ultra-violet photoemission spectroscopy (UPS) measurements on surfaces of blend films with varying donor to acceptor ratios[32] and extrapolating to a bulk heterojunction. The energetics was also estimated by simulations,[33–36] however these rely on reliable input parameters and do not substitute direct experimental measurements. Scanning Kelvin Probe Microscopy has been also applied for the study of energetic alignment in organic photovoltaic systems,[37–39] however this method suffers from limitations due to tip size convolution effects, complex sample preparation and doesn't offer information about the density of states (DOS) of the materials. Among the various experimental techniques offering information about the energy levels of materials, it is widely accepted that the most accurate and reliable method is photoemission spectroscopy. While these measurements enable a detailed investigation of the vacuum level position and the electronic structure of organic semiconducting materials,[40] their probing depth is limited to the top few nanometres of the sample, making them extremely surface sensitive.[41] Consequently, the investigation of energetic structure and alignment throughout thick layers of organic devices using this technique has been, so far, limited to thermally evaporated small molecules.[42–44] Such experiments involve consecutive steps of material deposition, few nanometres at a time, followed by spectroscopic characterisation in situ, typically referred to as 'layer by layer' investigation.

A 'layer by layer' approach is not compatible with solution-processed materials. The active layer of solution-processed devices is deposited in a single step, for example by spin-coating, resulting in a layer of tens to hundreds of nanometres.[45,46] This suggests that the only possible route for obtaining bulk information for solution-processed devices is by depth profiling. Molecular depth profiles yield surface sensitive information as a function of depth. They are achieved by a succession of etching steps and surface sensitive investigations, where the latter is dominantly performed by x-ray photoemission spectroscopy (XPS) or secondary ion mass spectrometry (SIMS). Monoatomic ion beam bombardment, typically consisting of Argon (Ar) ions, applied for many years as the main etching technique, is an effective etching tool, but is known to introduce significant surface damage, changing the chemistry of the sample, especially for organic materials,[47,48] making it unsuitable for application in combination with UPS. The recent introduction of argon gas cluster ion beams (GCIB)[49] for application in SIMS enabled an essentially damage-free etching, even for organic materials, with a sub-nm resolution by retaining molecular information.[50–57] In these sources, argon gas clusters are formed as Ar atoms coalesce following a supersonic expansion, resulting in clusters with a tuneable cluster size in the range of a few 1000 atoms/cluster and several eV per atom.[58] Originally employed in SIMS depth profiling, GCIB sources have already found application in combination with XPS,[59,60] however only a handful of reports exist of their combination with UPS. Specifically, a comparison of UPS surface measurements performed using the 'layer by layer' method with UPS depth profiling using GCIB confirmed an essentially damage-free removal of thermally evaporated organic molecules.[61,62] The combination of UPS with GCIB, however, has not been applied before for solution-processed material systems, such as the active layers of organic solar cells. Herein we demonstrate that ultra-violet photoemission spectroscopy depth profiling using a GCIB etching source allows the accurate determination of the energetic landscape of solution-processed bi-layer and bulk heterojunction active layers. First, we utilise the widely investigated model material system poly[3-hexylthiophene] (P3HT) and C61-butyric acid methyl ester (PC$_{60}$BM) (Chemical structures shown in **Suppl. Fig. 1)** and demonstrate that damage-free UPS depth profiling provides valuable information not only on the position of the donor and acceptor energy levels, but also on the formation of interfacial dipoles due to charge transfer, band bending and most surprisingly compositional information on a higher resolution that what can be obtained by XPS depth profiling. Next we apply our technique to four high efficiency systems, namely poly[(5,6-difluoro-2,1,3-benzothiadiazol-4,7-diyl)-alt-(3,3'''-di(2-octyldodecyl)-2,2',5',2'',5'',2'''-quaterthiophen-5,5'''-diyl)] (PffBT4T-2OD):[6,6]-Phenyl-C71-butyric acid methyl ester (PC$_{70}$BM), poly[(2,6-(4,8-bis(5-(2-ethylhexyl)thiophen-2-yl)-benzo[1,2-b:4,5-b']dithiophene))-alt-(5,5-(1',3'-di-2-thienyl-5',7'-bis(2-ethylhexyl)benzo[1',2'-c:4',5'-c']dithiophene-4,8-dione)] (PBDB-T):NCBDT, poly[(2,6-(4,8-bis(5-(2-ethylhexyl)thiophen-2-yl)-benzo[1,2-b:4,5-b']dithiophene))-alt-(5,5-(1',3'-bis(4-fluorothiophen-2-yl)-5',7'-bis(2-ethylhexyl)benzo[1',2'-c:4',5'-c']dithiophene-4,8-dione (PFBDB-T):{(2Z)-2-[(8-{(E)-[1-(dicyanomethylidene)-3-oxo-1,3-dihydro-2H-inden-2-ylidene]methyl}-6,6,12,12-tetraoctyl-6,12-dihydrothieno[3,2-b]thieno[2'',3'':4',5']thieno[2',3':5,6]-s-indaceno[2,1-d]thiophen-2-yl)methylidene]-3-oxo-2,3-dihydro-1H-inden-1-ylidene}propanedinitrile (C8-ITIC) and 2,2'-[(3,3''',3'''',4'-tetraoctyl[2,2':5',2'':5'',2''':5''',2''''-quinquethiophene]-5,5''''-diyl)bis[(Z)-methylidyne(3-ethyl-4-oxo-5,2-thiazolidinediylidene)]] bis-propanedinitrile (DRCN5T):PC$_{70}$BM and demonstrate its broad applicability to various material systems. We demonstrate that the photovoltaic gaps measured by UPS depth profiling are in close agreement with the measured CT state energies, which is not the case for photovoltaic gaps determined by measurements of the individual components of the blend. This observation demonstrates that interfacial effects that are present only when the donor and acceptor are blended strongly influence the electronic alignment at their interface and the corresponding accessible photovoltaic gap. This highlights the need for energetic characterisation of the bulk heterojunction by UPS depth profiling or other methods which can correctly determine the photovoltaic gap, rather than traditional characterisation of the constituent materials which are assumed to result in vacuum level alignment when blended, especially for novel material systems.

Finally, we apply our technique to a pristine and aged BHJ of DRCN5T:PC$_{70}$BM and poly({4,8-bis[(2-ethylhexyl)oxy]benzo[1,2-b:4,5-b']dithiophene-2,6-diyl}{3-fluoro-2-[(2-ethylhexyl)carbonyl]thieno[3,4-b]thiophenediyl}) (PTB7): PC$_{70}$BM (all chemical structures are shown in **Suppl. Fig. 2)** and demonstrate how UPS depth profiling is capable to probe the evolution of energetic levels upon environmental degradation, not currently possible with any other available method. We find that compositional variations in the BHJ result in formation of local band bending in the bulk of the active layer in the case of the small molecule DRCN5T system, which cannot be probed by performing UPS measurements only at the surface of the degraded samples.

## Results

**Monoatomic versus cluster etching modes and the UPS depth profiling method**

The efficacy of the ultra-violet photoemission depth profiling methodology relies on the ability to perform essentially damage-free etching, such that the electronic structure of the active layer materials remains unaltered by the etching steps, allowing for the acquisition of a meaningful UPS spectrum at any depth. Monoatomic Ar ion bombardment is a well-established etching tool that is particularly suitable for controlled removal of metals or other non-organic materials. While it is commonly used in combination also with organic materials (especially in the case of XPS depth profiling),[63–65] the high kinetic energy of the ions results in significant damage to such materials,[66–69] with recent results demonstrating that the damage introduced by monoatomic Ar ion bombardment extends far deeper into the bulk of organic layers than the XPS probing depth.[70] To circumvent this, our methodology is based on the use of an Ar GCIB source for etching, the working principle of which is schematically depicted in **Fig. 1a**. Ar clusters are formed by directing expanding argon gas into a nozzle followed by multiple skimmers. The clusters are then charged by a filament, size selected by magnetic lenses and focused by electrodes. A more detailed description of cluster and monoatomic ion beam formation is given in **Suppl. Note 1** and in references.[51,71–74] Both empirical measurements and molecular dynamic simulations of GCIB etching demonstrate the advantages of GCIB over conventional monoatomic ion sources.[75,76] Monoatomic etching results in a deep penetration into the sample and a large degree of mixing/damage. The etching yield is far lower than that of a cluster, hence the sample would have to be bombarded numerous times to eject a comparable amount of material, increasing the degree of accumulated damage even further.[77] Cluster etching results in the formation of a shallow crater with a low degree of surface damage both for inorganic and organic materials including polymers and biological samples.[78,79] The key differences in the impact between monoatomic and cluster etching modes are schematically summarised in **Fig. 1b**. The succession of etching steps with GCIB and UPS is shown for clarity in **Fig. 1c**.

To validate the feasibility of the new methodology, we investigated the effect of both monoatomic and cluster etching on the composition and electronic structure of several organic photovoltaic materials. (**Suppl. Fig. 3 and 4**) For example, a comparison of representative S2p (for P3HT) and O1s (for $PC_{60}BM$) XPS spectra and HOMO region UPS measurements for both materials after a 15 s etching step with either monoatomic or cluster source is shown in **Suppl. Fig. 3**. Two monoatomic ion energies (1000 eV and 2000 eV) and four cluster energies (2000 eV, 4000 eV, 6000 eV and 8000 eV) were selected to identify the optimum etching conditions. Regardless of ion energy, following a monoatomic etching step, a significant broadening of the full width half maximum (FWHM) of both the O1s and S2p XPS spectra are observed. This is accompanied by the complete disappearance of the valence band features for either material. In contrast, neither the composition nor the electronic structure of the materials is altered upon cluster etching, which is in excellent agreement with previous investigations.[61] We note that in the case of O1s spectra of $PC_{60}BM$, the differences in the relative ratio of the two oxygen species does not originate from etching induced damage, but rather from elimination of surface contamination that is more pronounced in the case of larger cluster energies as they result in increased etch rates. Further information about damage characterisation of other photovoltaic systems can be found in **Suppl. Note 2** and **Suppl. Fig. 4**.

To further highlight the necessity for GCIB as the etching source, we carried out a depth profile experiment on a $P3HT:PC_{60}BM$ film using the standard monoatomic Ar ion etching source, shown in **Suppl. Fig. 5**. We observe that after only a few etching steps, no identifiable valence band structure can be measured as the organic materials within the blend are heavily damaged by the monoatomic etching. A direct comparison to a profile obtained using a GCIB source (which will be discussed in more detail in the next sections) shows that GCIB etching preserves the electronic structure throughout the layer, allowing to extract meaningful information about both the electronic structure and composition as a function of depth.

Based on these results, we have chosen 4000 eV clusters for all further investigations since they were found to induce no damage to the electronic structure of the materials, ensure depth resolution on the sub-nm scale and result in reasonable etching times for layer thicknesses typical for photovoltaic devices.

**Combined XPS and UPS depth profiling of a $P3HT:PC_{60}BM$ bulk heterojunction (BHJ)**

The joint acquisition of both XPS and UPS spectra as a function of depth allows the extraction of both compositional and energetic depth profiles. We note that the technique does not provide lateral resolution, as both XPS and UPS signals are collected from an area of ~ 0.7 $mm^2$. The XPS depth profile allows us to follow the evolution of the atomic percentages of the individual atoms throughout the BHJ layer with a ~ 10 nm resolution determined by the probing depth of XPS. These atomic percentages allow the calculation of the donor (or acceptor) material percentages using their chemical formulae. The UPS depth profile contains valuable information about the energetic landscape in the photovoltaic active layer, namely the positions of the vacuum level, Fermi level and HOMOs for the top ~ 1-2 nm of the film surface at every depth. We note that HOMO level positions were determined by fitting the low binding energy edge of the valence band for the corresponding materials.[31] Furthermore, it is worth mentioning, that this determination results in the measurement of the ionization potential of the organic layer, which is the correct quantity for photovoltaic gap determination as highlighted by Brédas and coworkders already a decade ago.[80] However, the use of HOMO and LUMO terms (rather than the ionisation potential, IP, and electron affinity, EA) is ubiquitous in literature, so in the following we carry on with this notation. Nevertheless, we stress that our determined HOMO and LUMO energies are not the purely one-electron based theoretical constructs, but rather the positions of the measured electronic state energies.

A representative 3D energetic map, resulting from a UPS depth profile of a 60 nm thick BHJ of $P3HT:PC_{60}BM$ spin-coated on ZnO is shown in **Fig 2a.** The depth scale is subdivided into three regions: the top few nm (surface), the bulk (10-50 nm) and the region at the ZnO interface (around 60 nm) in order to highlight regions of particular interest. The valence band region (up to 6 eV below the Fermi level) is shown in **Fig 2b**. The evolution of the secondary electron onset (representing changes in the work function (WF)) and the valence band regions can be clearly observed and require detailed analysis.

In order to extract the positions of the HOMO levels of both the donor and acceptor materials in the bulk, it is necessary to disentangle their relative contributions to the measured UPS spectrum of the blend. Yan *et. al*.[81] proposed that the valence band spectrum of a blend could be fitted by a linear combination of the valence band spectra of the individual constituent materials. Following this approach allows us to identify the positions of the HOMO features of both the donor and the acceptor materials with respect to the Fermi level as well as to extract the relative contribution of each material to the overall measured UPS spectrum (As schematically shown in **Suppl. Fig 6**). The latter is directly related to the material percentages in each depth and can be compared with the results of XPS depth profiling (**Fig. 2c**), calculated by the measured sulphur/carbon (S/C) atomic ratio (as described in **Suppl. Note 3**).

Representative slices of the energetic map are shown in **Fig. 2c** surrounding the compositional profiles comparison and are denoted with 1 to 7. In each panel, the original UPS measurements (as shown in **Error! Reference source not found.a and 2b**) are plotted as orange circles, the fitted contributions of P3HT, $PC_{60}BM$ and ZnO (derived from the individual spectra as described in **Suppl. Note 3**) are shown in blue, red and green, respectively, whereas the sum of all three is shown as a purple line, referred to from now on as the overall fit. The top surface (depth point 1) UPS spectrum consists of only P3HT HOMO features in agreement with previous measurements.[63,82] The contribution of $PC_{60}BM$ HOMO features gradually increases in depth points 2 (3 nm) and 3 (7 nm), corresponding to increasing amounts of $PC_{60}BM$ in these depths. This increase is in agreement both with the results of XPS depth profiling performed here by cluster etching and previous reports obtained by monoatomic etching[63] and angle resolved XPS.[83] The key difference between the XPS and UPS derived composition profiles originates from the different probing depth of each of the two techniques. As the XPS probing depth is ~5 times larger than that of UPS, the former results in a smoothening effect of areas where stark variations in composition occur. For example, the compositional information at the blend surface includes also the contribution of deeper layers, such that the measured S/C ratio corresponds to a lower % of donor at the surface than it should be. The contribution of adventitious carbon from the surface further enhances this mismatch. Using the UPS depth profile to extract compositional information significantly increases the depth resolution due to the enhanced surface sensitivity of UPS and eliminates the effects of adventitious carbon. In the bulk of the film, no strong variations in composition are present and both methods result in a plateau (depth point 5). At depth point 6 (59 nm) near the interface with ZnO, a significant contribution of the ZnO valence band can be observed. Finally, only the contribution of ZnO to the UPS spectrum can be seen at depth point 7 (61 nm).

Etching a sample that consists of two different constituent materials (donor and acceptor) raises the question of preferential etching. Preferential etching occurs when one of the materials is etched at a faster rate than the other, resulting in the accumulation of the latter such that the measured profile no longer represents the actual sample. Three observations confirm that preferential etching does not occur in our case. Firstly, the existence of a plateau region in the bulk of the BHJ as measured by both the XPS and UPS profiling shows that the surface composition as probed by UPS is representative of the somewhat deeper measurement of composition by XPS that has yet to be affected by etching. Secondly, preferential etching would result in gradual accumulation of the material, which is etched more slowly, yet we observe a flat plateau in the bulk of the sample, rather than a slowly increasing contribution of one of the components. Finally, atomic force microscopy (AFM) measurements confirm that etching results in smooth surfaces throughout the entire sample, up until the ZnO interface. Representative height and adhesion micrographs collected after etching of 1 nm, 20 nm and 60 nm are shown in **Suppl. Fig. 7**. The surface roughness stays approximately constant (2.0 ± 0.3 nm) throughout the entire $P3HT:PC_{60}BM$ layer, increasing to 3.1 nm only upon reaching the underlying ZnO (at the depth of 60 nm). The adhesion, which is a measure for material contrast, shows no correlation with the height within the bulk, and changes only upon reaching the ZnO interface.

**Combined XPS and UPS depth profiling of a $P3HT/PC_{60}BM$ bilayer (BL)**

To investigate further the capabilities of UPS depth profiling as a tool for high-resolution compositional profiling, we performed a similar experiment on a bilayer of $P3HT/PC_{60}BM$. A representative 3D map, resulting from a UPS depth profile of a BL consisting of 55 nm of P3HT on top of 110 nm of $PC_{60}BM$ spin-coated onto ZnO is shown in **Fig. 3a and 3b**. In this case the change of the electronic structure from that of P3HT to $PC_{60}BM$ and also to that of ZnO can be clearly seen.

A comparison of the compositional profiles obtained by XPS and UPS depth profiles (**Fig. 3c**) confirms our previous observations. While the overall shape of the two profiles is in agreement, the sharpness of interfaces obtained *via* the UPS depth profile is superior to that of XPS. A detailed look at the representative slices 1-7 provides further insights. Firstly, the results show that GCIB etching does not result in any re-deposition or implantation of overlayer material into the bottom layers. The consistency of the electronic structure of each material at each depth demonstrates that even while etching through thick layers, no damage is introduced and/or accumulated during the experiment. Furthermore, while at depth point 1, the XPS and UPS compositional information are in excellent agreement, at depth point 2 (44 nm) the UPS spectrum shows 90% contribution of P3HT, while the XPS data predicts only 65%. This is a direct result of the larger probing depth of XPS, where signal originating from the deeper $PC_{60}BM$ layer results in an artificial reduction of the calculated P3HT percentage, significantly hindering the possible depth resolution of this technique. We note that even with UPS depth profiling we observe a mixed interfacial area (marked in grey) of about 7 nm (defined as the depth required for reducing the P3HT contribution from 90% to 10%). The lack of an absolutely abrupt interface in this material system can be explained by the diffusion of $PC_{60}BM$ molecules into the P3HT layer above as has been described previously.[84,85]

**Energetic landscape of $P3HT:PC_{60}BM$ and $P3HT/PC_{60}BM$ and of other high efficiency photovoltaic systems**

While the relative contributions of each of the constituent components provides us with compositional information on a nm length scale, the changes in the secondary photoemission onset and the positions of the low binding energy edge of the corresponding valence band spectra allow us to extract the energetic landscape of the photovoltaic system. These measurements provide us with the positions of the vacuum level and the HOMO w.r.t. the Fermi Level of each material as a function of depth. We note that directly extracting the LUMO positions would require performing inverse photoemission spectroscopy measurements, which due to low signal to noise ratio require acquisition at multiple spots, exhibit a large experimental error and often induce damage to organic layers.[86,87] For these reasons the LUMO positions were approximated by subtracting the optical gap of each material from the measured HOMO position and are only presented as an estimate.[88] The corresponding optical gaps are summarized in **Suppl. Table 1** and **Suppl. Fig. 8**. We note that this approximation does not account for differences between transport and optical transport levels due to the effects of both the exciton binding energy and the broadening of the optical gap as discussed in detail by Menke *et al*,[89] however as will be discussed later still results in a good estimate for the photovoltaic gaps. The resulting energetic landscapes for the BHJ and BL measurements are shown in **Fig. 4a** and **b**, respectively. We emphasise that despite being the most investigated organic photovoltaic system to date, the discrepancies in the representation of the energetic level diagram of this system persist with a large scatter of energy values reported in literature.[90–92]

Following the evolution of the energetic landscape of the BHJ reveals two interfacial effects. Starting at the interface with ZnO, a dipole can be observed through which the work function of ZnO (3.6 eV) is increased to 4.0 eV due to charge transfer to the $PC_{60}BM$, or possibly due to polarization effects,[29] in agreement with previous reports.[63,93] This interfacial region is followed by the bulk of the blend layer in which the energy levels remain constant. We note that due to the presence of $PC_{60}BM$ in the blend, the work function is limited by the LUMO of $PC_{60}BM$[94] and cannot be below 4.0 eV. At the surface of the blend, a thin layer of P3HT is formed, so in the top 1-2 nm, this restriction is lifted and the work function is reduced, resulting in a bending of the P3HT energy levels at the surface. This bending of the P3HT energy levels is in excellent agreement with previously reported measurements performed by varying the P3HT:$PC_{60}BM$ ratio, in which an increasing amount of $PC_{60}BM$ resulted in a shift in WF and P3HT HOMO level with an overall band bending on the order of 0.5 eV.[82]

The interaction between P3HT and $PC_{60}BM$ is evident in the energetic landscape of the BL. While a similar dipole at the ZnO/$PC_{60}BM$ interface is observed as in the BHJ case, a second dipole can be seen at the P3HT/$PC_{60}BM$ interface. This interfacial dipole corresponds to charge transfer from the P3HT onto the $PC_{60}BM$ as also suggested in previous reports.[95] Our measurements reveal a 0.3 eV dipole, in excellent agreement with measurements performed on P3HT/$PC_{60}BM$ bilayers formed by in situ electrospray deposition (0.35 eV dipole)[96] and measurements on P3HT/$C_{60}$ (0.3-0.4 eV dipole).[97] We note that this charge transfer from P3HT results in gradual band bending: the region directly near the interface exhibits a smaller energetic distance between the Fermi level and the HOMO, corresponding to a lower density of electrons in the proximity of the interface. Moving further away from the interface (towards the surface) results in an increase in this energetic distance and a higher density of electrons, consistent with electron transfer taking place at the interfacial region only. We note that we cannot rule out that very small amounts of $PC_{60}BM$ have diffused into the P3HT layer,[84,85] influencing the band bending near the P3HT/$PC_{60}BM$ interface.[82] The small upward bending of P3HT at the very surface can be attributed to the fabrication method including exposure to water and oxygen of the bilayer device.[98]

Both systems result in a similar estimated photovoltaic gap (calculated as the difference between the measured HOMO of the donor and the estimated LUMO of the acceptor) of 1.03±0.04 eV and 1.16±0.02 eV for the BHJ and BL, respectively. We note, that these values are estimations due to combined transport and optical measurements and that the errors listed above are of statistical nature, *i.e.* the result of averaging individual measurements collected throughout the entire film thickness. The overall experimental error is approximately 0.1-0.15 eV, originating from both measurement and fitting errors. While literature values for the photovoltaic gap, even for this extensively investigated material system vary from 0.9 eV to 1.4 eV,[99–102] our measurements are in good agreement with the measured 0.9 eV photovoltaic gap measured directly by Park *et al*[96] and in close agreement with the CT state energy measured via luminescence spectroscopy.[18,103] To demonstrate the applicability of our technique beyond the model of P3HT:$PC_{60}BM$, we chose four high performance photovoltaic systems, with either fullerene or non-fullerene acceptors and both polymer and small molecule donors. The first system, PffBT4T-2OD:$PC_{70}BM$, has been investigated by us and others and resulted in power conversion efficiencies (PCE) approaching 11%.[104,105] The two systems based on non-fullerene acceptors, PBDB-T:NCBDT and PFBDB-T:C8-ITIC result in impressive PCEs of 12-13%.[7,8] Last but not least, the small molecule DRCN5T:$PC_{70}BM$ system results in a PCE of ~9 %.[106] The 3D energetic maps of these photovoltaic systems are shown in **Suppl. Fig. 9**, and the resulting energetic landscapes including the estimated LUMO levels are shown in **Fig. 4 (c)-(f).** The chemical structures of the materials are shown in **Suppl. Fig. 2**. Unlike P3HT:$PC_{60}BM$, no strong surface segregation of the either of the components is observed. The extracted estimated photovoltaic gap for both PffBT4T-2OD:$PC_{70}BM$ and DRCN5T:$PC_{70}BM$ is 1.31±0.02 eV. In the case of non-fullerene acceptor systems, these values are 1.48±0.03 eV and 1.61±0.02 eV for PBDB-T:NCBDT and PFBDB-T:C8-ITIC, respectively. Both values are in good agreement with reported CT energy estimates for these material systems (1.54 and 1.53 eV, respectively), obtained by characterising the overlap of the external quantum efficiency and the electroluminescence of the blend.[8,107] As mentioned above, the errors listed for the photovoltaic gaps are of statistical nature, demonstrating that the photovoltaic gap remains highly constant with almost no deviation throughout the bulk heterojunction.

While the estimated photovoltaic gaps include uncertainties due to the use of optical gaps, our method provides accurate measurements of the HOMO-HOMO offsets for all investigated material systems. The importance of these energetic offsets has been recently demonstrated by Qian *et al*.[17] It has been shown that low energetic offsets are one of two prerequisites for low voltage losses and high power conversion efficiencies in organic solar cells. While these offsets could be estimated by, for example by measurements on each blend component separately, such an estimate will omit taking into account interfacial interaction effects (such as formation of dipoles, see **Fig. 4b**) and the evolution of these offsets across the device. Our method accurately measures these offsets throughout the active layer and can be easily applied to characterise new material systems in order to identify candidates that satisfy the condition of a low energy offset. In our case, we observe high HOMO-HOMO offsets for all systems that employ fullerene derivatives and remarkable low offsets for the two non-fullerene derivatives, the latter being critical for their high photovoltaic performance. The estimated photovoltaic gaps and HOMO-HOMO offsets are summarised in **Table 1** with open-circuit voltages of corresponding devices provided for reference.

**A comparison of estimated photovoltaic gaps from different approaches with device $V_{OC}$s**

The practice of calculating the photovoltaic gap based on the energy levels of the individual blend components is excessively common in literature. While in some cases energy values for both components are extracted from photoemission spectroscopy measurements, oftentimes different methods for energy level measurement for each of the blend components are used introducing additional errors. To evaluate the agreement between the photovoltaic gaps calculated from measurements performed on the individual blend components and those obtained by our UPS depth profiling method on BHJ active layers to the reported CT energy values and the corresponding $V_{OC}$s, both are shown in **Fig. 5** for the six photovoltaic systems investigated here. We note that reliable values for the CT energies are available from literature for five out of the six systems studied here.[8,103,107–109] The results clearly show that photovoltaic gaps obtained from individual components are in poor agreement with the measured CT state energies and cannot be correlated to the measured $V_{OC}$s. In some cases, the extracted photovoltaic gap is even lower than the measured $V_{OC}$. On the contrary, the photovoltaic gaps extracted from UPS depth profiling measurements on BHJ layers are in a good agreement with the measured CT energies and can be clearly correlated to the $V_{OC}$s with a certain offset. We note that this offset, and thus the difference between the CT state energy and the $V_{OC}$ is not necessarily a constant value and depends on the radiative and non-radiative losses in each particular system, both resulting from charge recombination.[110] While all solar cells loose some voltage due to radiative recombination, non-radiative recombination can be geminate or non-geminate and includes recombination through traps, structural defects, triplet states and Auger recombination.[110] Correspondingly and in agreement with our results, it is not expected that all the photovoltaic gaps fall on a single line with a slope of 1, but rather show some variation along that line. We also want to point out that the photovoltaic gaps calculated from the UPS depth profiling data include an error introduced by the use of the optical gap for the calculation of the LUMO of the acceptor in each system, but even with this error taken into account, they represent a far more accurate picture of the energetic alignment at the donor/acceptor interface than the one obtained from individual measurements of the blend components. It is also interesting to note that the agreement with the measured CT state energies suggests that the energetic alignment as obtained via UPS depth profiling is representative of the origin of the CT state energy, and can be used to study new photovoltaic systems.

The observation that UPS depth profiling of BHJ layers correctly probes the photovoltaic gaps, whereas measurement on separate materials fails, suggests that significant ground state electronic interaction takes place between the blend components in many cases. This is not unexpected, as for example in the case of the P3HT/PCBM bilayer, an interfacial dipole is formed upon contact between the donor and the acceptor (**Fig. 4b**). The nature of this electronic communication will depend on the exact donor and acceptor, but also on the microstructure of the blend. As a result, performing the measurements by UPS depth profiling on blends of identical microstructure to the photovoltaic devices is invaluable.

**Evolution of the energetic landscape upon degradation**

Despite tremendous advances in the performance of organic photovoltaic devices, environmental degradation remains the Achilles heel of devices based on organic materials. Development of future mitigation strategies necessitates that the effects of environmental factors on the device performance are thoroughly investigated and understood.

Changes in the energetic landscape of the devices upon environmental degradation would play a crucial role determining their photovoltaic performance, as all fundamental processes are affected by these changes. While existing methods cannot probe these variations, UPS depth profiling can be applied to photovoltaic systems at any point in their lifetime. To demonstrate the broad applicability of our method to the study of degradation mechanisms of organic photovoltaic systems, we investigated the evolution of the energetic landscape of an efficient small molecule:fullerene system DRCN5T:PC$_{70}$BM and that of a polymer:fullerene system PTB7:PC$_{70}$BM. **Fig. 6** shows the energetic landscape and the compositional profile of a pristine and a degraded BHJ of these materials, where the degradation conditions were exposure to 1 sun illumination in dry air for 12h.

In the case of DRCN5T:PC$_{70}$BM, the pristine BHJ energy levels do not show any strong surface or interface effects, similar to the other high performance systems described earlier (**Fig. 4**). However, upon degradation, significant changes to the energetic landscape can be observed. Overall, the Fermi level position for both the donor and the acceptor shifts closer to the HOMO, suggesting p-doping by oxygen as has been previously observed for other photovoltaic systems.[111,112] However, while the surface measurements show a relatively minor change in the energy levels, a far stronger change occurs at a depth of about 10 nm below the surface. Interestingly, the compositional profiles reveal that this region shows an inversion in the donor:acceptor composition: while the surface is rich in the donor molecule, at around 10 nm below the surface the fullerene acceptor constitutes the majority of the overall composition at approximately 60%. These results demonstrate that the changes in the energy levels upon environmental degradation may not occur uniformly throughout the BHJ, but rather vary depending on the compositional profile of the blend. In particular, we observe that a stronger p-doping by oxygen occurs at a region rich with fullerene molecules, as the energy levels at this region shift upwards more than for those rich with the donor molecules. It is interesting to compare these results to those of PTB7:PC$_{70}$BM, in which the compositional profile below the surface PTB7 rich layer remains constant with no fullerene rich inversion region. Upon degradation the energy levels show a uniform p-type doping effect, as expected from oxygen, confirming that the local variation observed for degraded blends of DRCN5T:PC$_{70}$BM is associated with their compositional variations. Such local effects in the bulk of the photovoltaic blend cannot be probed by other methods, but are critically important as they influence processes such as charge generation and transport.

## Discussion

The development of UPS depth profiling opens a vast array of opportunities for the study of organic photovoltaic systems. It is a valuable tool that could help to better understand the organic-organic interface alignment, specifically the characterisation of the HOMO-HOMO offset, but also that of the formation of the CT state. The method allows to estimate the CT state energy which can be very challenging for some material systems. By being able to predict the CT state energy, we are also able to predict the potential $V_{OC}$ of the device without the need for empirical optimisation.

Furthermore, the extraction of both high-resolution compositional information and energy level evolution from a single measurement has fascinating implications. For example, ternary organic photovoltaic systems have recently drawn the attention of the photovoltaic community due to impressive efficiencies surpassing 11%.[13] Much remains unknown about the energetic landscape and vertical compositional profiles in such systems, however it has been recently proposed that controlled ternary structures where sequential deposition from solution of the binary donor:acceptor system follows the deposition of the second acceptor can significantly enhance performance.[113] Alternatively, it has also been suggested that some ternary blends result in a spontaneous segregation of the third component to the surface of the blend resulting in an increased open-circuit voltage and improved charge extraction.[114] The application of our methodology to these ternary material systems will allow the accurate determination of both their energetic and compositional profiles.

The ability to cross organic/organic and organic/inorganic interfaces and accurately determine the energetic alignment makes our methodology of great interest to the study of complete photovoltaic devices (including charge extraction layers) as well as tandem organic photovoltaics. Organic tandem cells have recently made a remarkable come-back, with record efficiencies of 15%[115] and 17.3%[2] demonstrated in close succession. UPS depth profiling can be performed on the entire stack of a tandem device to extract the complete energetic landscape of such a system with high vertical resolution. It is important to mention that the technique can also be applied to devices after operation, allowing for the study of the temporal evolution of the energetic landscape upon aging, which can be of crucial importance, such as shown with our results in **Fig. 6**.

We note that our methodology is not limited to photovoltaic material systems. The technique is directly applicable to study energetic landscapes of any multi-layered device, including photodetectors and light-emitting diodes. Another possible application is the study of highly desired solution-processed doping of organic materials. Extensive research has led to the demonstration of both n-type and p-type doping with remarkable applications in both organic-light-emitting diodes and field-effect transistors.[116] However, accurately determining the distribution of the dopant in the host matrix is very difficult, especially in the case of surface doping.[117] UPS depth profiling will allow the simultaneous determination of the doping depth evolution and its effect on the energetic landscape within the layer.

Novel film structures, such as sequentially deposited donor–acceptor layers,[107] can be also investigated in great detail with our technique. Finally, the ability to extract compositional profiles from electronic structure measurements is in itself intriguing. Let us consider a mixed material system in which the two components are comprised from the same elements making compositional profiling challenging. If the two materials show dissimilar electronic structure, UPS depth profiling will be able to successfully distinguish them, whereas traditional methods such as XPS depth profiling would struggle or fail entirely.

To summarise, a novel UPS depth profiling technique where UPS measurements are combined with essentially damage free GCIB etching has been developed and demonstrated on a range of photovoltaic material systems. The new methodology accurately determines the energetic landscape of both bilayer and bulk heterojunction systems, allowing an estimation of the photovoltaic gap and the accurate determination of HOMO-HOMO offsets for each material system. Additionally, the method offers compositional profile information on a much larger depth resolution than that offered by for example XPS depth profiling. The method has vast applications to the fields of organic electronics and can be applied to the study of energetic landscapes of any type of multi-layered device. Finally, as UPS depth profiling can be performed at any point in the device lifetime, it offers valuable insights also into the temporal evolution of the electronic and compositional profiles after device operation. This includes determining the effects of environmental degradation on the device energetic landscapes, necessary for the development of more stable devices.

## Experimental Procedures

**Materials.** P3HT and PC$_{60}$BM were obtained from Sigma Aldrich, PffBT4T-2OD from 1-Material, PBDB-T and PC$_{70}$BM from Ossila (M1002). Zinc acetate dihydrate as a precursor to ZnO was bought from Sigma-Aldrich. All materials were used as received without further purification. NCBDT, PFBDB-T and C8-ITIC were synthesized using procedures reported elsewhere.[7,8] Chloroform, DCM and Chlorobenzene were purchased from Sigma Aldrich.

**Film preparation.** *P3HT and PC$_{60}$BM*: First, a ZnO solution was prepared with the sol-gel method[118] and substrates with a pre-patterned ITO slide on glass (PsiOTec Ltd., UK) were cleaned in an ultrasonic bath with acetone and isopropanol for 5 min each and O$_2$ plasma cleaned for 10 min at 0.4 mbar. A ZnO film was then spin-coated on the cleaned substrates. For bulk heterojunction films, pre-weighted P3HT and PC$_{60}$BM were brought into a glovebox (< 1 p.p.m. O$_2$ and H$_2$O), solved in chlorobenzene (16 mg/ml) and stirred on a hotplate (70 °C) over night. The solutions were mixed for a donor to acceptor mass ratio of 1:0.8 and spin-coated on ZnO/ITO/Glass substrates using a 0.45 µm PTFE filter and with a spin speed of 1000 r.p.m. (60 s). Bilayer films were prepared using the float-casting technique. Films of P3HT (8 mg/ml) and PC$_{60}$BM (40 mg/ml) were separately spin-coated on PEDOT:PSS and ZnO, respectively, at the very same way as described above for the bulk heterojunction films. The substrates were then transferred to ambient atmosphere, the P3HT films were immersed in DI water and floating onto a deionized water surface. The free floating P3HT film on water was then picked up by the substrate with PC$_{60}$BM on top. Films were then placed under vacuum ($10^{-1}$ mbar) for 30 min to ensure proper film formation and water evaporation. The films were not annealed in order to prevent unnecessary fullerene intermixing.

*PffBT4T-2OD:PC$_{70}$BM*: Following the previously described cleaning procedure of ITO and ZnO deposition, the substrates were transferred to a N$_2$-filled glovebox (<2 ppm of O$_2$, < 2 ppm of H$_2$O) and got preheated on a hot plate at 90 °C. The active layer was spin-coated from a chlorobenzene:dichlorobenzene (1:1) solution of mixed PffBT4T-2OD:PC$_{70}$BM in a weight ratio of 16:19.2 mg mL$^{-1}$ kept at 90 °C. The additive 1,8-diiodoctane was added 2 h prior to deposition as 3 vol %. A plastic chuck was used for the spin-coating step to avoid cooling of the substrates during film deposition. The samples were then annealed for 5 min at 80 °C.

*PBDB-T:NCBDT:* The PBDB-T:NCBDT BHJ was deposited on glass/ITO/PEDOT:PSS substrates. ITO coated glass substrates were cleaned sequentially by deionized water, acetone and isopropyl alcohol under ultrasonication for 10 min each. The subsequent PEDOT:PSS layer was spin-coated at 5000 RPM for 45 s, and then baked at 150 ºC for 20 min in ambient atmosphere. Next, the active layer was spin-coated from donor (5 mg/mL) and acceptor (4 mg/mL) in chloroform solution at 1400 rpm for 20 s. The active layers were placed in a glass Petri dish for solvent vapor annealing.

*PFBDB-T:C8-ITIC*: For the PFBDB-T:C8-ITIC BHJ layers, the ITO cleaning and ZnO deposition was carried out the same way as for P3HT:PC$_{60}$BM. The blend solution were dissolved in chlorobenzene (CB) at a ratio of 1:1.25, and a concentration of 20 mg/ml, and were stirred overnight at 50 °C. The blend solution were spin coated onto the ZnO coated ITO substrate (2000 rpm, 1 min).

*DRCN5T:PC70BM:* For the DRCN5T:PCBM BHJ layers, ITO-patterned glass substrates were cleaned and ZnO was deposited following the aforementioned procedures. DRCN5T and PC$_{70}$BM were dissolved in chloroform and mixed in overall concentrations of 15mg/mL and 12mg/mL before spin-coating at 1700 rpm for 20s. The films were annealed for 10 min at 120°C (thermal annealing) and additionally for 1 min in a petri-dish rich in chloroform vapor (solvent annealing). To degrade the BHJ, the sample was exposed to dry air under 1 sun illumination for 12h prior to UPS depth profiling measurements.

**UPS/XPS measurements.** The samples were transferred to an ultrahigh vacuum chamber (ESCALAB 250Xi), with a base pressure of $2*10^{-10}$ mbar, for UPS/XPS measurements. UPS measurements were performed using a double-differentially pumped He gas discharge lamp emitting He I radiation (hv=21.22 eV) with a pass energy of 2 eV and a bias of -5 V in order to ensure secondary electron onset detection. The UPS spectra are shown as a function of the binding energy with respect to the Fermi Energy. The energy edge of the valence band is used to determine the HOMO level with respect to the Fermi level and the secondary electron onset with combination of the latter are used to determine the vacuum level with respect to the Fermi level, denoted also as the work function. XPS measurements were carried out using an XR6 monochromated Al Kα X-ray source (hv = 1486.6 eV) with a 650 µm spot size.

## Acknowledgements

**General:** The authors would like to kindly thank Prof. U. Bunz for providing access to film fabrication facilities and Prof. A. Pucci for access to AFM.

**Funding:** V. L. and Y.V. thank the Juniorprofessor Program of the Baden-Württemberg Ministry of Science, Research and Art for funding. J. Z. thanks the China Scholarship Council for a PhD scholarship (No. 201503170255). This project has received funding from the European Research Council (ERC) under the European Union's Horizon 2020 research and innovation programme (ERC Grant Agreement n° 714067, ENERGYMAPS).

## Author Contributions

Y. V. conceived, guided and supervised the project. V.L. performed and analysed the UPS depth profiling experiments. A. W., J. Z., Y. C., M. H, and R. H. F. performed and supervised the synthesis and layer fabrication for the high efficiency photovoltaic materials. V.L. and Y.V jointly wrote the manuscript, which was edited by all other co-authors.

## Declaration of Interests

The authors declare no competing financial interests.

# Tables

**Table 1 | Estimated photovoltaic gaps ($E_{PV}$), open-circuit voltages ($V_{OC}$) and HOMO-HOMO offsets of BHJs of P3HT:PC$_{60}$BM and the high efficiency PffBT4T-2OD:PC$_{70}$BM, PBDB-T:NCBDT and PFBDB-T:C8-ITIC systems.** Note that the listed errors are statistical errors, resulting from measurements of multiple sampling depths. Each individual measurement has an error of 0.15 eV originating from measurement and fitting errors.

| Acceptor type | Material system | $E_{PV}$ (eV) | $V_{OC}$ (V) | HOMO-HOMO Offset (eV) |
|---|---|---|---|---|
| Fullerene | P3HT:PC$_{60}$BM | 1.03±0.04 | 0.6 | 0.97±0.04 |
|  | PffBT4T-2OD:PC$_{70}$BM | 1.31±0.02 | 0.75[104] | 0.54±0.02 |
|  | DRCN5T:PC$_{70}$BM | 1.31±0.02 | 0.93 | 0.53±0.02 |
|  | PTB7:PC$_{70}$BM | 1.17±0.03 | 0.77 | 0.68±0.02 |
| Non-fullerene | PBDB-T:NCBDT | 1.48±0.03 | 0.84[7] | 0.02±0.02 |
|  | PFBDB-T:C8-ITIC | 1.61±0.02 | 0.93[8] | 0.12±0.02 |

# Figures

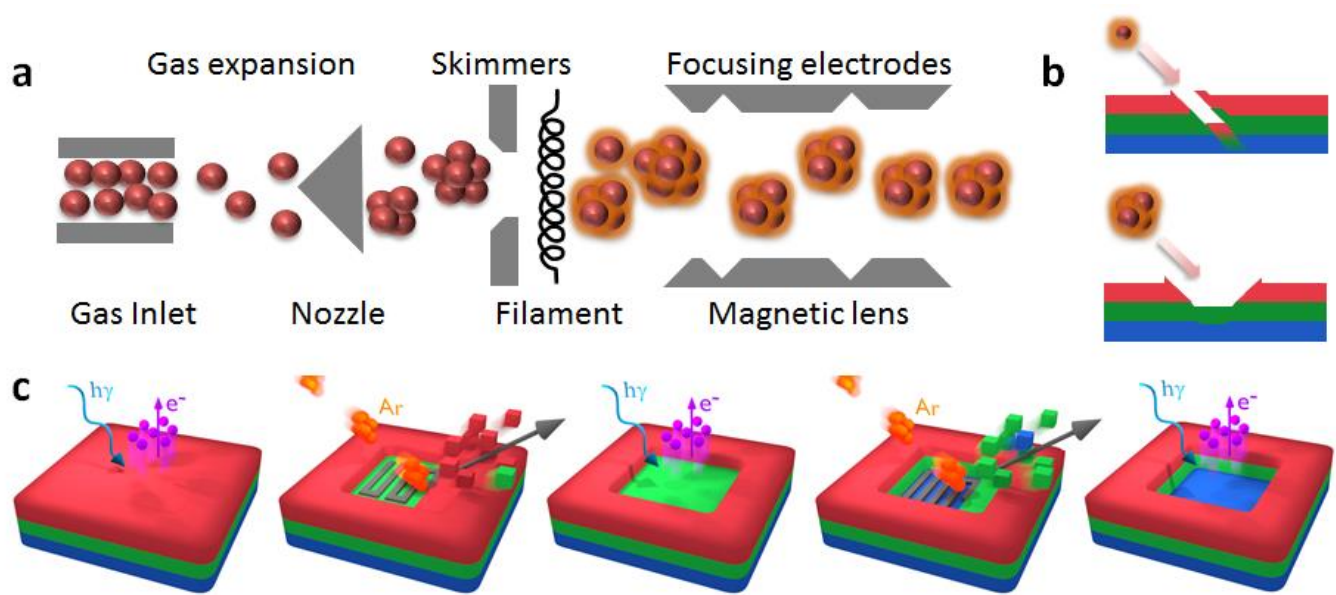

**Fig. 1 | Argon beam formation, impact difference of monoatomic and cluster modes and schematics of our etching technique. (a)** Functional principle of an argon gas cluster ion gun: The Argon gas is expanded through a gas expansion region. After reaching a nozzle, clusters are formed. Skimmers with differentially pumped vacuum regions ensure stable cluster formation. By flying through a filament, the clusters are bombarded by electrons and become charged. Cluster size selection and beam focusing are provided by magnetic lenses and focusing electrodes, respectively. **(b)** Schematics of impact differences of monoatomic and cluster beams on a thin multiple-layer film. **(c)** Schematic representation of the succession of surface sensitive investigations with UPS and etching steps with Argon GCIB. UPS measurements are represented with a blue photon, indicating the incident UV light, and with emitted electrons. Ar etching is represented with cluster balls flying towards the surface, resulting in material being removed, exposing the next layer via rastering, denoted by a grey line.

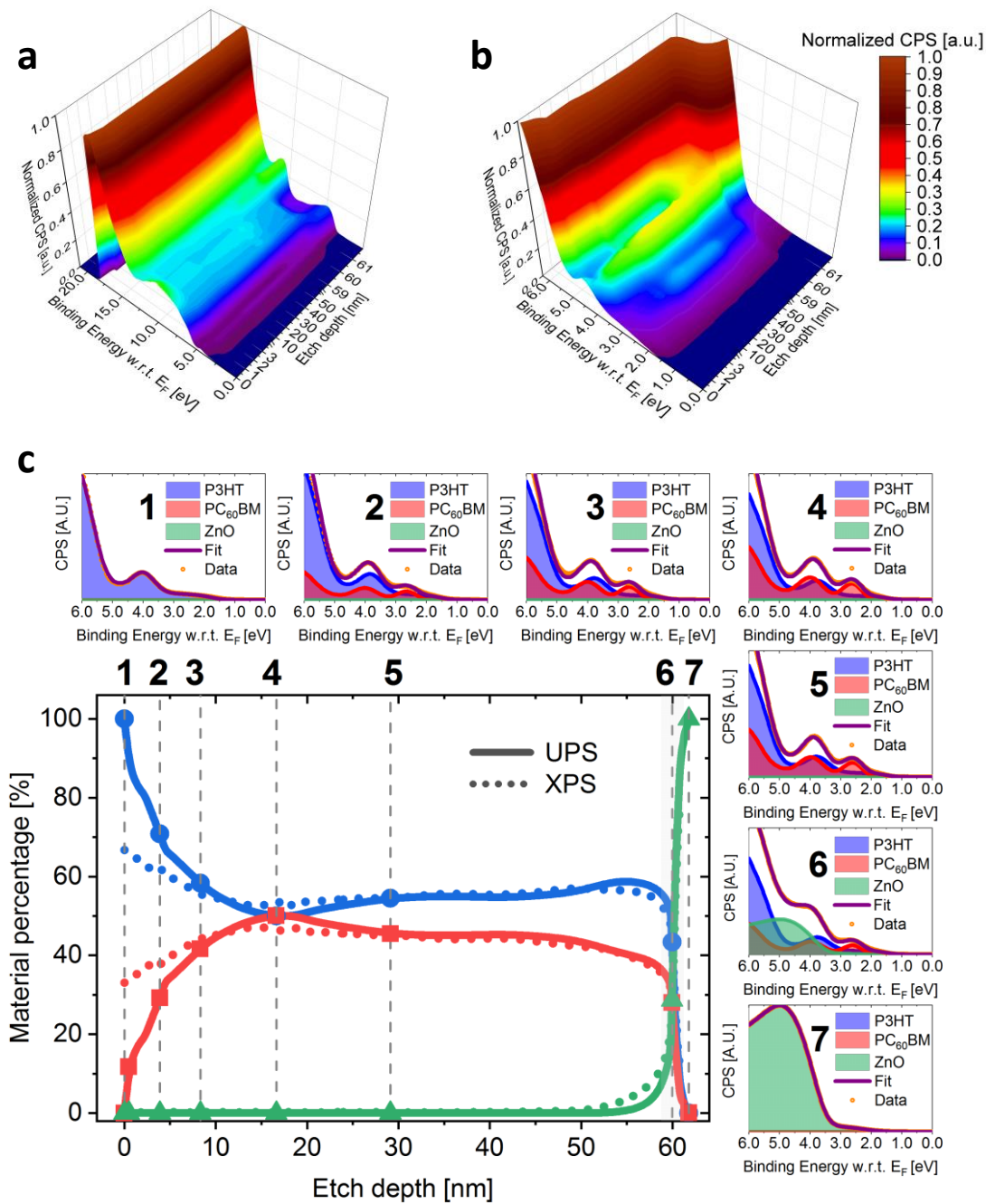

**Fig. 2 | Energetic map of a spin-coated, 60 nm thick P3HT:PC$_{60}$BM bulk heterojunction on ZnO and the extracted material percentage as a function of etch depth from UPS and XPS depth profiles with representative individual UPS measurements and fittings.** The full UPS depth profile is shown in **(a)**, including the secondary electron onset. The valence band region near the Fermi level is showed in **(b)**, highlighting the HOMO region throughout the layer. The spectra are plotted with respect to the Fermi Energy, denoted as E$_F$. **(c)** The material percentage data from XPS measurements was calculated from the measured S/C atomic ratios at each depth. The UPS data was extracted from our novel fitting method. Seven representative valence band spectra and corresponding fits are shown on the top and right side of the atomic percentage graph and correspond to the marked points in the main graph. The single fits for P3HT, PC$_{60}$BM and ZnO are denoted with blue, red and green colours, respectively. The sum of the 3 individual contributions (purple) is in excellent agreement with the measured data (orange dots).

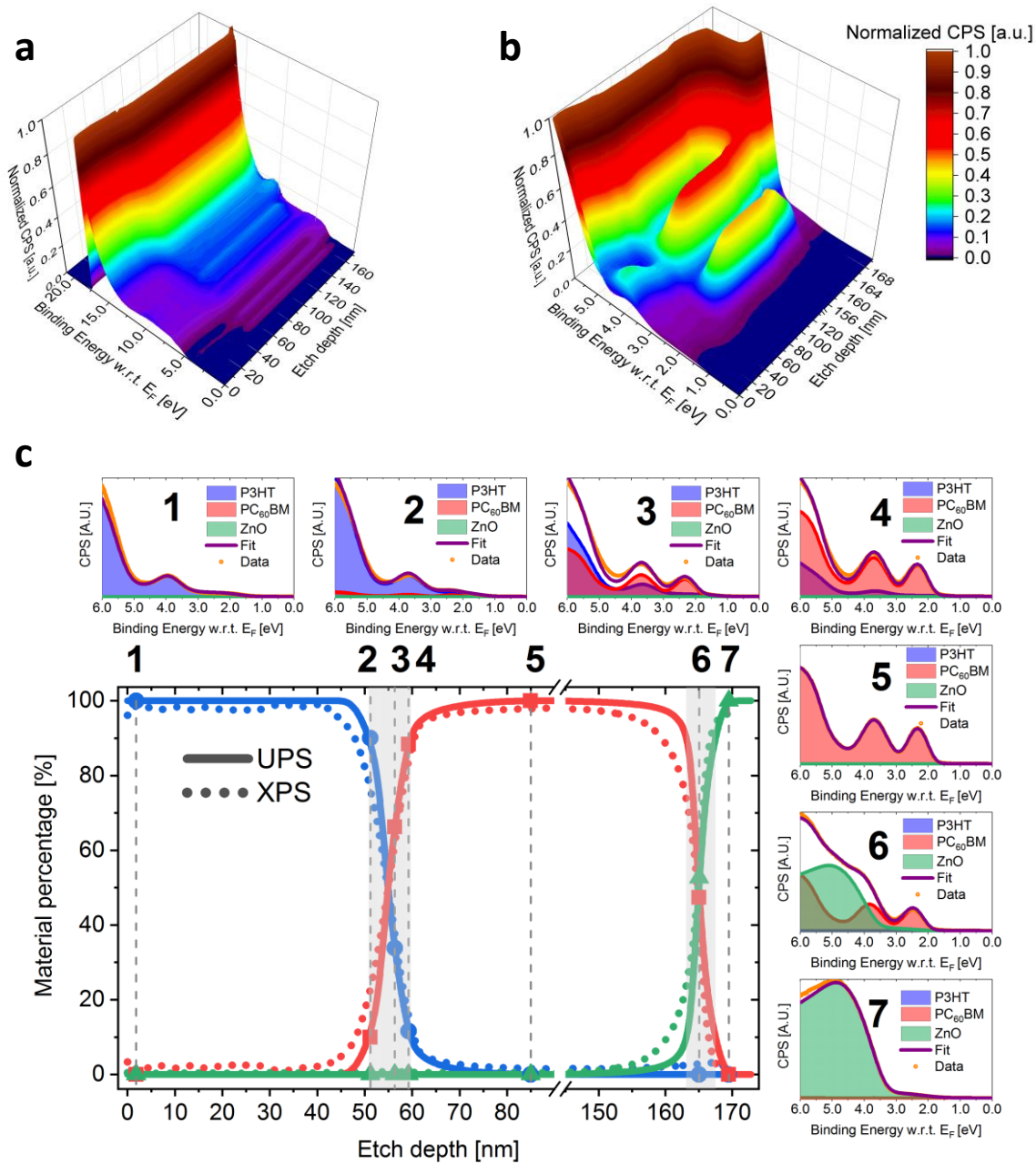

**Fig. 3 | Energetic map of a spin-coated, bi-layered P3HT (55 nm) and PC$_{60}$BM (110 nm) film on ZnO and material percentage as a function of etch depth from UPS and XPS depth profiles with representative individual UPS measurements and fittings.** The full UPS depth profile is shown in **(a)**, including the secondary electron onset. The valence band region near the Fermi level is showed in **(b)**, highlighting the HOMO region throughout the layer. The spectra are plotted with respect to the Fermi Energy, denoted as E$_F$. **(c)** The material percentage data from XPS measurements was calculated from the measured S/C atomic ratios at each depth. The UPS data was extracted from our novel fitting method. Seven representative valence band spectra and corresponding fits are shown on the top and right side of the atomic percentage graph and correspond to the marked points in the main graph. The single fits for P3HT, PC60BM and ZnO are denoted with blue, red and green colours, respectively. The sum of the 3 individual contributions (purple) is in excellent agreement with the measured data (orange dots).

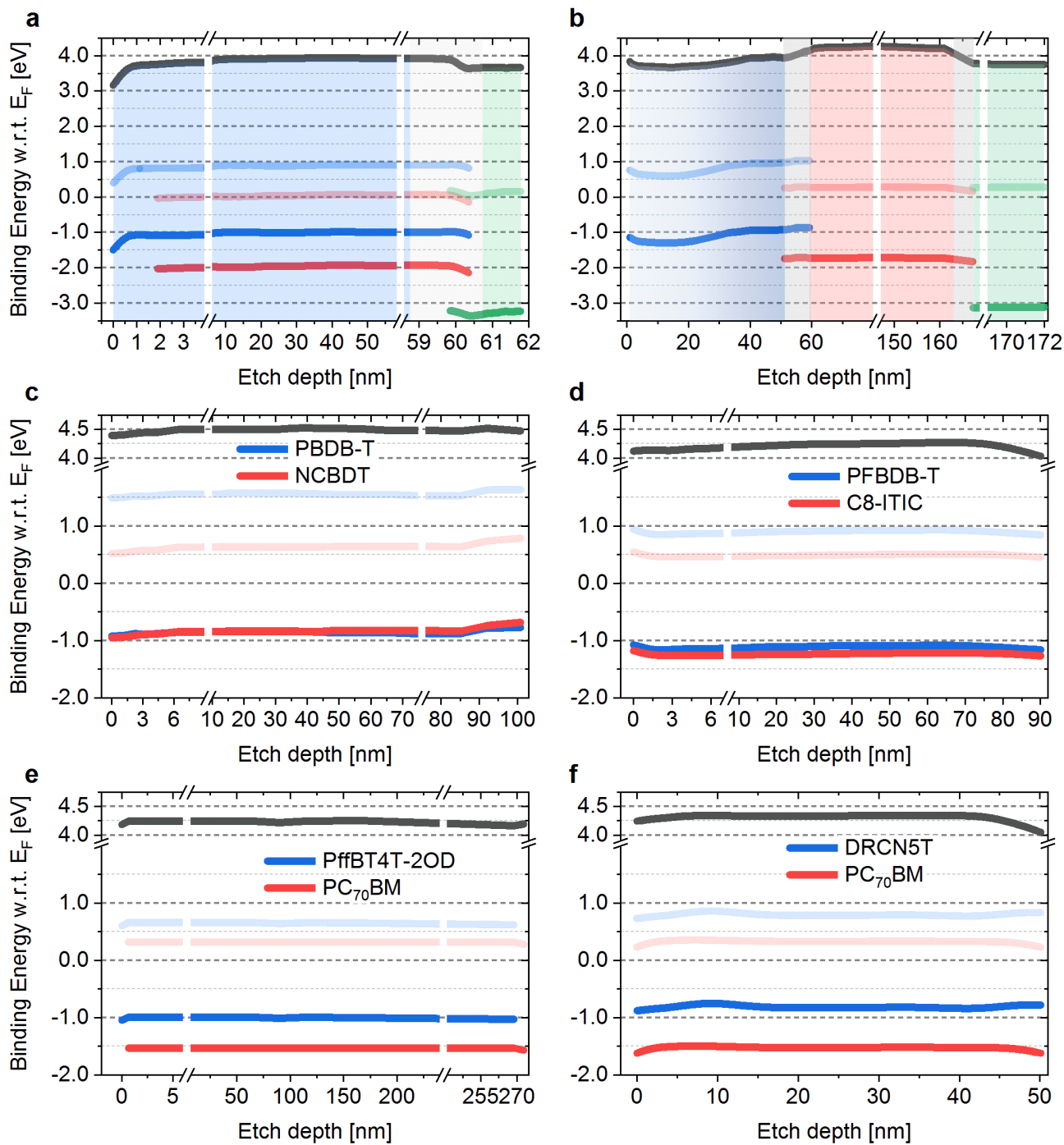

**Fig. 4 | Energetic landscapes of P3HT:PC$_{60}$BM and P3HT/PC$_{60}$BM and other high efficiency photovoltaic systems.** Vacuum level (black), measured valence and estimated conduction band positions with respect to the Fermi Level of the donor (blue), and acceptor (red) materials for the investigated **(a)** bulk heterojunction P3HT:PC$_{60}$BM, **(b)** bilayer P3HT/PC$_{60}$BM, **(c)** PBDB-T:NCBDT, **(d)** PFBDB-T:C8-ITIC, **(e)** PffBT4T-2OD:PC$_{70}$BM and **(f)** DRCN5T:PC$_{70}$BM films. In the case of the P3HT films, also the underlying ZnO is shown in green. The chemical structures of the materials are provided in **Suppl. Fig. 1 and 2**. Optical gaps are given in **Suppl. Table 1**.

Table 1 | **Estimated photovoltaic gaps ($E_{PV}$), open-circuit voltages ($V_{OC}$) and HOMO-HOMO offsets of BHJs of P3HT:PC$_{60}$BM and the high efficiency PffBT4T-2OD:PC$_{70}$BM, PBDB-T:NCBDT and PFBDB-T:C8-ITIC systems.** Note that the listed errors are statistical errors, resulting from measurements of multiple sampling depths. Each individual measurement has an error of 0.15 eV originating from measurement and fitting errors.

| Acceptor type | Material system | $E_{PV}$ (eV) | $V_{OC}$ (V) | HOMO-HOMO Offset (eV) |
|---|---|---|---|---|
| Fullerene | P3HT:PC$_{60}$BM | 1.03±0.04 | 0.6 | 0.97±0.04 |
|  | PffBT4T-2OD:PC$_{70}$BM | 1.31±0.02 | 0.75[104] | 0.54±0.02 |
|  | DRCN5T:PC$_{70}$BM | 1.31±0.02 | 0.93 | 0.53±0.02 |
|  | PTB7:PC$_{70}$BM | 1.17±0.03 | 0.77 | 0.68±0.02 |
| Non-fullerene | PBDB-T:NCBDT | 1.48±0.03 | 0.84[7] | 0.02±0.02 |
|  | PFBDB-T:C8-ITIC | 1.61±0.02 | 0.93[8] | 0.12±0.02 |

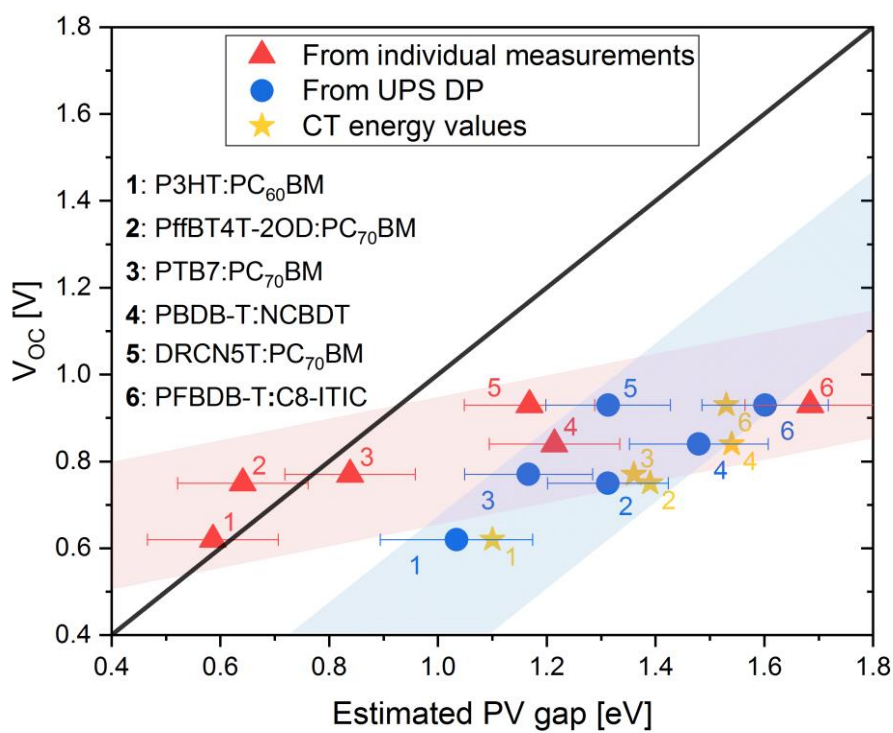

**Fig. 5 | Empirical relationship between estimated photovoltaic gaps from different approaches and the $V_{OC}$ of the corresponding devices.** Photovoltaic gaps are estimated from individual measurements (red triangles) and UPS depth profile measurements (blue dots). Literature values for CT energies of systems with very similar film morphology are shown as yellow stars. The investigated systems are denoted with numbers from 1 to 6. All six investigated systems are denoted with numbers from 1 to 6. Blue and red semi-transparent bars with a thickness of ~0.07 eV are guides for the eye and represent fits of the corresponding estimated gaps.

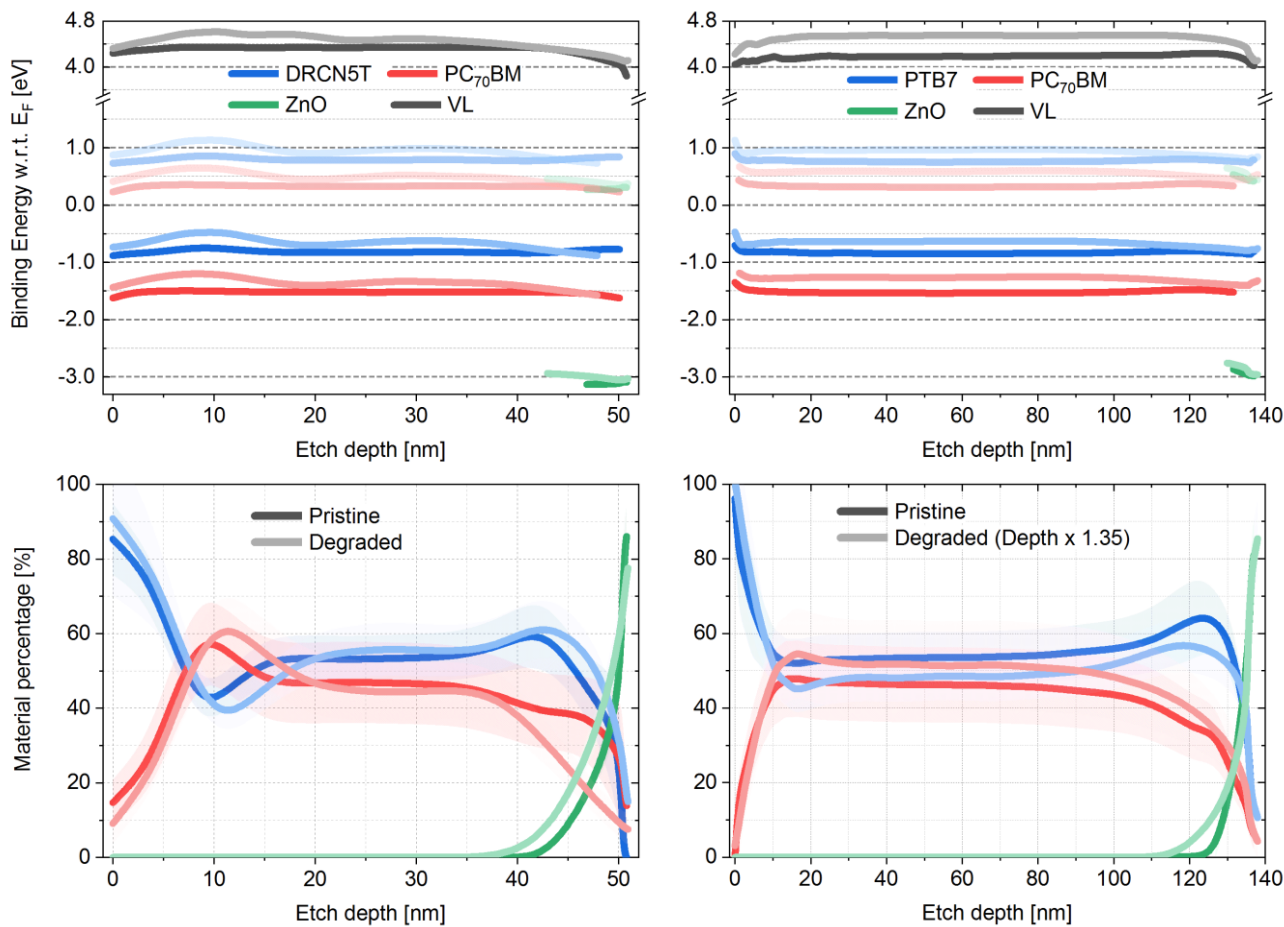

**Fig. 6 | Change of the energetic landscape (top) and material percentage (bottom) landscape of the DRCN5T:PC$_{70}$BM (left) and PTB7: PC$_{70}$BM (right) photovoltaic systems.** Vacuum level (black), valence band and estimated conduction band (semi-transparent) positions with respect to the Fermi level of the donor (blue), acceptor (red) and the underlying ZnO (green). Pristine data is plotted in dark, while degraded data in light colours. The chemical structures of the materials are provided in **Suppl. Fig. 2**. Optical gaps are given in **Suppl. Table 1**.